\documentclass[twocolumn ]{aastex701}
\usepackage{amsmath}
\usepackage{bm}
\usepackage{booktabs}
\usepackage{rotating}
\begin{document}

\title{Escaped White Dwarf Candidates from  Open Clusters}

\correspondingauthor{Huahui Yan}

\author[orcid=0009-0005-8565-0858]{Huahui Yan}
\affiliation{School of Physics and Astronomy, Sun Yat-sen University, DaXue Road, Zhuhai 519082, GuangDong, People’s Republic of China.}
\affiliation{CSST Science Center for the Guangdong-Hongkong-Macau Greater Bay Area, Sun Yat-sen University, DaXue Road, Zhuhai 519082, GuangDong, People’s Republic of China}
\email[show]{yanhh8@mail2.sysu.edu.cn}

\author[0000-0002-4591-1903]{David R. Miller}
\affiliation{Department of Physics and Astronomy, University of British Columbia, Vancouver, BC V6T 1Z1, Canada}
\email[]{}  

%
\author{Jingkun Zhao}
\affiliation{National Astronomical Observatories, Chinese Academy of Sciences, Beijing 100101,  People’s Republic of China}
\email[]{}  
\author{Jincheng Guo}
\affiliation{Department of Scientific Research, Beijing Planetarium, Xizhimenwai Street, Beijing 100044,  People’s Republic of China}
\email[]{}  
\author{Chengyuan Li}
\affiliation{School of Physics and Astronomy, Sun Yat-sen University, DaXue Road, Zhuhai 519082, GuangDong, People’s Republic of China.}
\affiliation{CSST Science Center for the Guangdong-Hongkong-Macau Greater Bay Area, Sun Yat-sen University, DaXue Road, Zhuhai 519082, GuangDong, People’s Republic of China}
\email[]{lichengy5@mail.sysu.edu.cn}
\begin{abstract}
Observations reveal a pronounced deficit of white dwarfs (WDs) in open clusters (OCs) relative to theoretical expectations, suggesting that a significant fraction of WDs may have escaped from their parent clusters after formation. In this work, we perform a systematic search for escaped WD candidates from OCs by back-tracing the motions of WDs and star clusters from Gaia DR3 catalogs. We identify 476 candidate WDs with kinematics consistent with having escaped from one of 175 OCs.  A control‑field Monte Carlo (MC) test yields a contamination rate of $87.6\% \pm 4.3\%$.
Given this high contamination rate, we filter the candidates by ensuring each WD's total age is consistent with its host cluster age, thereby establishing a more reliable follow-up sample of 109 stars. The excluded candidates with anomalous ages are more likely field interlopers or products of accelerated binary evolution. Among these, the low-mass regime exhibits a clear excess over expected field binary merger rates, whereas the high-mass regime remains broadly consistent with binary population synthesis predictions. Finally, comparing escaped WDs with cluster properties, absolute escape counts appear limited by Gaia's distance‑dependent incompleteness. The normalized escape fraction shows little dependence on cluster age, possibly favouring WD loss near formation over gradual evaporation, but is strongly anticorrelated with cluster mass, plausibly because deeper potential wells of massive clusters retain more of their WDs.

\end{abstract}


\section{Introduction} 

White dwarfs (WDs) represent the final evolutionary state of the vast majority ($\sim97\%$) of stars in the Galaxy \citep{Woosley2015, Lauffer2018}, formed as compact remnants of progenitors with initial masses up to $\sim8$--$10\,M_{\odot}$ \citep{Doherty2015}. Unable to ignite further fusion reactions of any kind, WDs are supported against gravitational collapse by electron degeneracy pressure. \citep{Chandrasekhar1931, Shapiro1983}. Their fundamental properties, including mass, effective temperature, and atmospheric composition, can be determined through spectroscopic and photometric analyses, providing essential constraints on stellar evolution theory \citep{Koester2009}. The observed mass distribution of WDs peaks near $0.608\,M_{\odot}$ \citep{Kepler2019}. However, a distinct population of WDs with masses below $0.3\,M_{\odot}$, referred to as extremely low-mass (ELM) WDs, cannot be produced through single-star evolution within a Hubble time. Their formation requires binary interaction, in which mass transfer strips the progenitor envelope before the core reaches the threshold for helium ignition. Observations confirm that the vast majority of ELM WDs are found in binary systems \citep{Kilic2011, Kilic2012, Brown2013, Brown2016}, establishing them as reliable tracers of binary evolution products among the WD population.

The combination of Gaia photometry and parallax allows WDs to be placed precisely on the Hertzsprung--Russell (HR) diagram, where they populate a faint and blue locus clearly separated from main-sequence (MS) stars. Their observed colors and absolute magnitudes are governed primarily by effective temperature and atmospheric composition, with secondary dependences on surface gravity, interior structure, and interstellar reddening. By comparing the observed HR diagram positions of WDs with theoretical
cooling models, allows the derivation of their masses and cooling ages. Successive Gaia data releases have enabled the assembly of progressively larger and more complete WD catalogs, from DR2 \citep{GaiaCollaboration2018A&A...616A..10G, Esteban2018MNRAS.480.4505J, Gentile2019, Pelisoli2019} to the substantially expanded DR3 samples \citep{GaiaCollaboration2021A&A...649A...6G, GentileFusillo2021, Esteban2023MNRAS.518.5106J}, greatly advancing statistical studies of WD populations across the Galaxy.

OCs serve as natural laboratories for stellar evolution,
as their member stars were formed within the same molecular cloud and therefore share nearly identical ages, metallicities, and kinematic properties. The
high-precision astrometry delivered by Gaia has greatly accelerated
OC discovery, since cluster identification relies on detecting stellar groups
with coherent positions, parallaxes, and proper motions. Applying a range of
clustering algorithms to successive Gaia data releases, numerous
studies have substantially expanded the known OC census
\citep{CastroGinard2018A&A...618A..59C, CastroGinard2019A&A...627A..35C,
	CastroGinard2020A&A...635A..45C, Liupang2019ApJS..245...32L,
	Cantat-Gaudin2018A&A...618A..93C, Hunt2021A&A...646A.104H, Hunt2023,
	Hunt2024}, uncovering a growing population of previously unrecognized
systems across the Galactic disk.

OCs are also among the most valuable settings for studying WD evolution.
Their well-determined ages directly constrain the progenitor lifetime of
each member WD, enabling a straightforward estimate of its initial mass.
This property makes OCs the primary benchmarks for empirically calibrating
the initial--final mass relation (IFMR), the fundamental mapping between a
star's birth mass and the mass of the WD it ultimately produces
\citep[e.g.,][]{Weidemann1977, Koester1981, Reimers1982, Weidemann1987, Weidemann2000, Williams2004, Kalirai2005, Dobbie2006, Williams2007,Kalirai2008, Rubin2008, Cummings2015, Cummings2016, Cummings2018, Richer2021, Prisegen2021, Marigo2020, Heyl2022, Miller2022, Prisegen2023, Miller2026}.
A key assumption underlying this calibration is that the WDs formed through
single-star evolution, which requires their total ages to be consistent with
the host cluster age. A WD whose astrometric parameters---position, proper motion, and parallax---are consistent with cluster membership, yet whose total age exceeds the cluster age, cannot have formed through single-star evolution. Analogous to the formation of ELM WDs, such objects most likely originated through binary interactions that accelerated the evolution of the progenitor. Nevertheless, it is important to acknowledge that such discrepancies may simply arise from coincidental associations, wherein a field WD fortuitously shares the kinematic properties of the cluster without being physically bound to it. Despite this potential contamination, age discrepancies remain a useful indicator for identifying objects that have likely undergone non-standard evolution. Building on this approach, \citet{Yan2026} carried out a systematic search for WD members in OCs and, by comparing total ages with cluster ages, identified a sample of candidates whose formation likely proceeded through binary evolutionary channels. In a follow-up study, \citet{Yan2025} reported the discovery of a rapidly rotating, strongly magnetized WD in a 35~Myr-old OC, with properties consistent with accelerated formation via binary evolution.

Observations have revealed a pronounced deficit of WDs in OCs relative to theoretical predictions \citep{Eggen1965, Tinsley1974,Richer1998, Kalirai2001, Williams2007, Richer2021}. A well-known example is the nearby, extensively studied Hyades cluster, where roughly three-quarters of the expected WD population remains undetected \citep{Weidemann1992}. Given that the detection limit is not a restricting factor for such proximate systems, sample incompleteness alone cannot account for this observed shortfall. The most natural interpretation is that a substantial fraction of WDs have escaped from their natal clusters. A widely invoked explanation for this ejection is that WDs receive natal velocity kicks at formation. During the AGB phase, non-spherical mass loss or dynamical encounters can impart velocities sufficient to eject newly formed WDs from the shallow gravitational potential of their host cluster \citep{Fellhauer2003, Heyl2007, Fregeau2009}. In the binary channel, \citet{Sandquist1998} demonstrated that asymmetric common-envelope (CE) ejection can deliver characteristic recoil velocities of 3--8\,km\,s$^{-1}$ to post-CE remnants, comparable to the escape velocities of typical OCs. Observational support for this picture was provided by \citet{Grondin2024}, who identified WD+MS binaries with proper motions consistent with cluster membership but exhibiting significant spatial offsets from the cluster center. Their analysis showed that for typical kick velocities, the resulting positional displacement is a more pronounced observable signature than the induced proper-motion deviation, making spatial offset the more sensitive diagnostic for identifying escaped WDs.

The persistent deficit of WDs in OCs has motivated systematic efforts to
recover the escaped population through dynamical back-tracing techniques.
\citet{Heyl2022} pioneered a particularly successful approach by tracing WD orbits backward in time using Gaia EDR3 astrometry and recovered two WDs with past trajectories consistent with ejection from the Pleiades. Applying a similar
kinematic retrieval strategy to a broader set of nearby clusters,
\citet{Miller2022, Miller2023} identified several escaped massive WDs,
including one with a mass of 1.317\,$M_{\odot}$, among the most massive WDs
known to have formed through single-star evolution. However, these studies
have focused on individual or small samples of nearby clusters, and a
comprehensive census of escaped WDs across the full OC population remains
lacking. In this work, we perform a systematic search for escaped WD
candidates from a large sample of OCs by kinematically back-tracing
Gaia DR3 WD and cluster catalogs, aiming to characterize the
escaped WD population on a statistical basis and to explore the dependence
of the WD escape fraction on cluster properties.

The remainder of this paper is organized as follows. In Section~\ref{sec2}, we describe the selection of our WD and OC samples. Section~\ref{sec3} details the dynamical back-tracing method used to identify escaped WDs and presents the main results. In Section ~\ref{sec4}, we estimate the contamination rate, define a high-confidence sample, and examining the dependence of the escaped WD recovery fraction on cluster parameters. We summarize our conclusions in Section~\ref{sec5}.


\section{Data Selection} \label{sec2}

\subsection{WD Sample}

The WD candidates used in this work are selected from the catalog of \citet{GentileFusillo2021}, which was constructed based on Gaia Early Data Release 3 (EDR3; \citealt{GaiaCollaboration2021A&A...649A...6G}) astrometry and photometry. This catalog employs a probabilistic classification scheme utilizing the position of sources in the Gaia color-magnitude diagram, and provides the probability of being a WD ($P_{\rm WD}$) for each candidate, along with derived atmospheric parameters including effective temperature, surface gravity, and mass. 

\begin{figure*}[htbp]
	\centering
	\includegraphics[width=\textwidth]{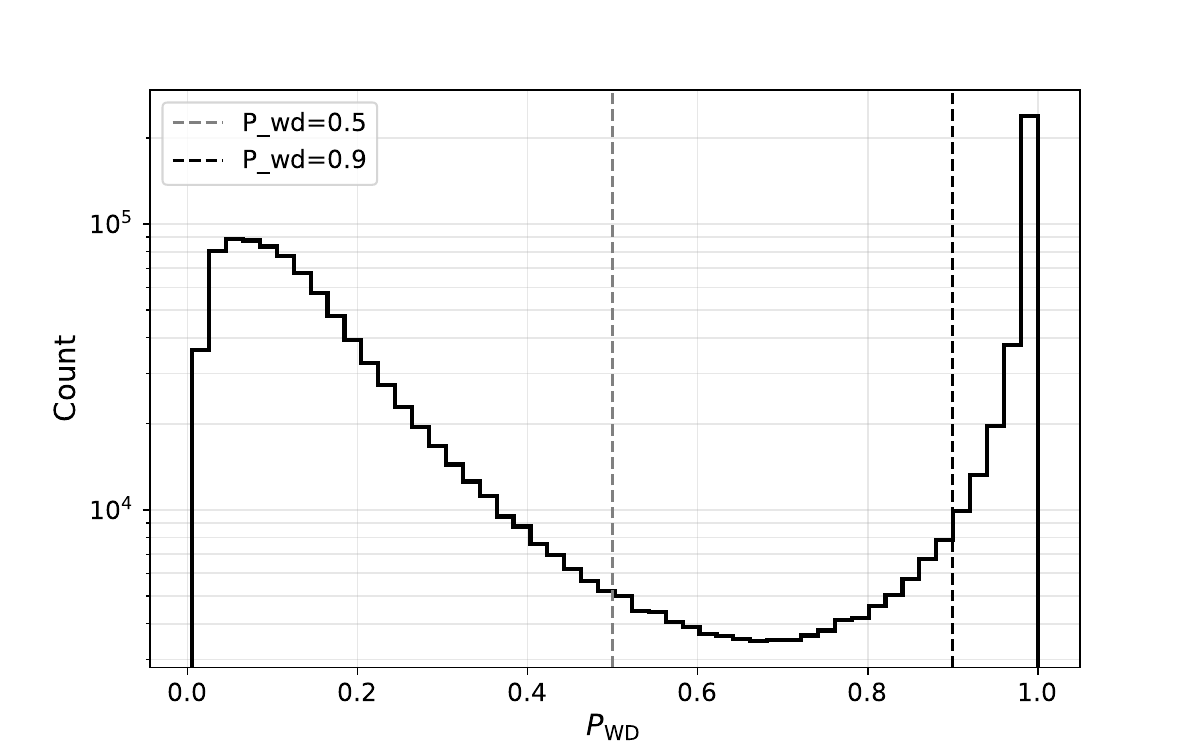}
	\caption{The logarithmic probability distribution of WDs ($P_{\rm WD}$) from \citet{GentileFusillo2021}. The grey dashed line represents the threshold of $P_{\rm WD} = 0.5$, and the black dashed line marks our adopted threshold of $P_{\rm WD} = 0.9$.}
	\label{fig:pwd_dist}
\end{figure*}

To establish a robust basis for our kinematic tracking, we set the 
	selection threshold for WD candidates from their probability parameter $P_{\rm 
		WD}$. As shown in Figure~\ref{fig:pwd_dist}, the $P_{\rm WD}$ values from the 
	\citet{GentileFusillo2021} catalog are clearly bimodal, with a low-probability 
	contaminant population well separated from a true-WD locus near $1.0$. We 
	therefore adopt a stringent cut of $P_{\rm WD} \ge 0.9$, placed within the gap 
	between the two peaks. Because the distribution is bimodal, this cut removes 
	most contaminants while reducing the sample only modestly, from $\sim$408,000 
	sources under a loose $P_{\rm WD} \ge 0.5$ cut to 319,251. This yields a clean 
	and reliable initial sample for the subsequent kinematic tracking.

\subsection{OC Sample}
The OC sample is drawn from the updated Milky Way star cluster catalog of \citet{Hunt2024} (hereafter H\&R24), which reanalyzes the earlier \citet{Hunt2023} cluster sample by incorporating photometric mass estimates and Jacobi radius calculations to distinguish gravitationally bound clusters from unbound moving groups. This classification identifies 5,647 of their 7,167 catalogued systems as bounds OCs. To further ensure sample reliability, H\&R24 applied two additional quality metrics: a Cluster Significance Test (CST) based on nearest-neighbor distributions, and a homogeneity parameter derived from a convolutional neural network trained on simulated clusters. Following their recommendations, we select clusters with CST $> 5\sigma$ and homogeneity $> 0.5$, and restrict our sample to OCs only, yielding a high-quality subset of 3,530 OCs. H\&R24 provides a comprehensive set of cluster parameters, including central coordinates ($\alpha$, $\delta$), parallax, proper motions ($\mu_{\alpha}$, $\mu_{\delta}$), radial velocity, distance, age, extinction ($A_V$), and cluster radius. We adopt these parameters in our search for escaped WDs from OCs. Considering the astrometric precision of Gaia, we further restrict our sample to clusters within 1.1 kpc of the Sun \citep{Prisegen2021}. These additional selection criteria yield a final sample of 752 OCs.

\section{IDENTIFYING ESCAPED WDs}\label{sec3}

\subsection{Method}
To identify WDs that have escaped from OCs, we propagate the positions of WD candidates backward in time over a duration corresponding to the cluster age, and select those whose past positions fall within prescribed distance criteria of the cluster center. The method follows a similar approach to that of \citet{Heyl2022,Miller2022,Miller2023}. For each WD candidate, we first calculate the relative velocity with respect to each cluster in the plane of the sky:
\begin{equation}
	\Delta v_{\rm 2D} = v_{\rm 2D} - v_{\rm cluster} + \frac{v_{\rm cluster} \cdot \boldsymbol{r}}{\boldsymbol{r} \cdot \boldsymbol{r}} \boldsymbol{r},
\end{equation}
where $v_{\rm 2D}$ is the velocity of the WD in the plane of the sky derived from Gaia proper motions, $v_{\rm cluster}$ is the cluster velocity, and $\boldsymbol{r}$ is the displacement vector of the WD relative to the Sun.  Since radial velocities are unavailable for most WD candidates, we assume zero radial velocity in this initial calculation. 

To determine whether a WD could have originated from a given cluster, we compute the distance between the WD and the cluster center as a function of time:
\begin{equation}
	d(t)^2 = [\boldsymbol{r} - \boldsymbol{r}_{\rm cluster} + t(v_{\rm 2D} - v_{\rm cluster}) + \hat{\boldsymbol{r}}\delta r]^2,
\end{equation}
where $t = 0$ corresponds to the present and $\delta r$ represents the radial displacement. We then solve for the time $t_{\rm min}$ at which the WD was closest to the cluster center:
\begin{equation}
	t_{\rm min} = \frac{\Delta \boldsymbol{r} \cdot \Delta v - (\Delta \boldsymbol{r} \cdot \hat{\boldsymbol{r}})(\Delta v \cdot \hat{\boldsymbol{r}})}{(\Delta v \cdot \hat{\boldsymbol{r}})^2 - (\Delta v)^2},
\end{equation}
where $\Delta \boldsymbol{r} = \boldsymbol{r} - \boldsymbol{r}_{\rm cluster}$ and $\Delta v = v_{\rm 2D} - v_{\rm cluster}$. This allows us to reconstruct the radial velocity as:
\begin{equation}
	v_r = -\frac{\hat{\boldsymbol{r}} \cdot (\Delta \boldsymbol{r} + t_{\rm min} \Delta v)}{t_{\rm min}},
\end{equation}
and thus obtain the three-dimensional velocity $\hat{v}_{\rm 3D} = v_{\rm 2D} + v_r \hat{\boldsymbol{r}}$ and the reconstructed three-dimensional velocity $\Delta \hat{v}_{\rm 3D} = \hat{v}_{\rm 3D} - v_{\rm cluster}$. A WD is considered an escaped candidate if it satisfies the following criteria: (1) the minimum approach distance $d_{\rm min} < 15$~pc, (2) the closest approach occurred in the past ($t_{\rm min} < 0$) and within the cluster's age, and (3) the reconstructed relative velocity satisfies $\Delta \hat{v}_{\rm 3D} < 5$~km\,s$^{-1}$.  The spatial constraint of $d_{\rm min} < 15$~pc accounts for both the typical tidal radius of OCs and the accumulation of astrometric uncertainties over the integration time. The kinematic constraint of $\Delta \hat{v}_{\rm 3D} < 5$~km\,s$^{-1}$ is consistent with the shallow gravitational potential wells of OCs. It accommodates typical cluster escape velocities combined with reasonable natal velocity kicks imparted during WD formation, while vigorously rejecting high-velocity background field stars \citep{Heyl2022,Binney2008gady.book.....B}.  
\subsection{Cluster Ages} \label{clusterage}

\begin{figure*}
	\centering
	\includegraphics[width=0.8\textwidth]{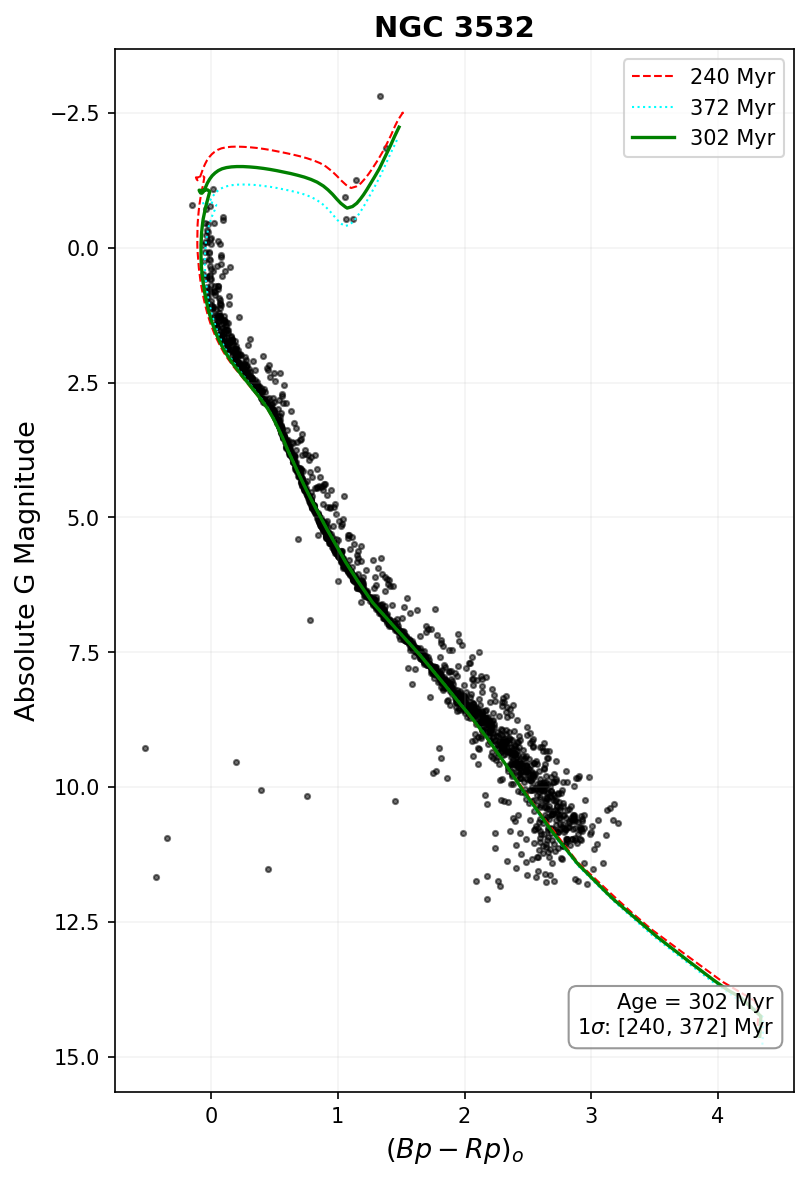}
	\caption{CMD of the NGC 3532  cluster. Black dots represent highly probable cluster members ($P > 50\%$) from the H\&R24 catalog. The green solid line indicates the best-fit PARSEC isochrone, while the red and cyan dashed lines represent the $1\sigma$ error boundaries.}
	\label{fig:example_fit}
\end{figure*}

To accurately identify escaped WDs from OCs, a reliable estimate of the cluster age is essential. We determine the ages of the OCs in our sample by fitting PARSEC 2.0 isochrones \citep{Bressan2012, Nguyen2022} to their color--magnitude diagrams (CMDs), following the methodology described in \citet{Miller2026}. Using the mean cluster age and reddening from H\&R24 as initial parameters, we adopt a fixed solar metallicity of Z = 0.015 and adjust the reddening $A_V$ in steps to achieve the best visual match to the central main sequence. The best-fit age is determined using a weighted $\chi^2$ statistic, where stars near the main-sequence turnoff (MSTO) receive the highest weight (60\%), giant stars contribute 30\%, and main-sequence stars closely aligned with the isochrone account for the remaining 10\%. The upper and lower $1\sigma$ uncertainty bounds are obtained by iterating the age in both directions from the best fit until the weighted $\chi^2$ increases by 30\%. While the $\chi^2$ values provide an initial guide for the fitting process,
the final adopted ages are determined through careful visual inspection of
the isochrone fit for 752 OCs, with particular emphasis on the agreement in
the MSTO region. An example isochrone fit for NGC 3532 is shown in Figure~\ref{fig:example_fit}, and the derived ages for all 752 OCs are listed in Table~\ref{tab:cluster_params}.

\begin{deluxetable*}{lrrrrrrrrrcc}
	\tablecaption{Parameters of the 752 open clusters analyzed in this work.\label{tab:cluster_params}}
	\tablewidth{0pt}
	\tablehead{
		\colhead{Name} & \colhead{Age$_{\rm best}$} & \colhead{Age$_{\rm lower}$} & \colhead{Age$_{\rm upper}$} & \colhead{$A_V$} & \colhead{R.A.} & \colhead{Decl.} & \colhead{$\mu_{\alpha}\cos\delta$} & \colhead{$\mu_{\delta}$} & \colhead{$\varpi$} & \colhead{$N_{\rm WD,exp}$} & \colhead{$N_{\rm esc}$} \\
		\colhead{} & \colhead{(Myr)} & \colhead{(Myr)} & \colhead{(Myr)} & \colhead{(mag)} & \colhead{(deg)} & \colhead{(deg)} & \colhead{(mas\,yr$^{-1}$)} & \colhead{(mas\,yr$^{-1}$)} & \colhead{(mas)} & \colhead{} & \colhead{}
	}
	\startdata
	ASCC\_101 & 289 & 238 & 368 & 0.099 & 288.345 & 36.360 & 0.967 & 1.281 & 2.510 & 8 & 10 \\
	ASCC\_105 & 63 & 59 & 93 & 0.200 & 295.472 & 27.386 & 1.370 & -1.618 & 1.776 & 2 & 0 \\
	ASCC\_106 & 121 & 97 & 151 & 1.218 & 295.792 & 1.563 & 0.810 & -2.493 & 1.327 & 2 & 0 \\
	ASCC\_107 & 18 & 15 & 24 & 1.350 & 297.152 & 21.980 & -0.197 & -5.122 & 1.115 & 0 & 0 \\
	ASCC\_11 & 286 & 204 & 316 & 0.600 & 53.050 & 44.841 & 0.905 & -3.072 & 1.171 & 18 & 0 \\
	\multicolumn{12}{c}{$\cdots$} \\
	vdBergh\_83 & 8 & 6 & 10 & 0.221 & 99.992 & -27.182 & -2.851 & 3.294 & 1.043 & 0 & 0 \\
	ASCC\_125 & 4 & 3 & 5 & 3.825 & 344.102 & 62.711 & -0.541 & -2.197 & 1.191 & 0 & 0 \\
	COIN-Gaia\_5 & 288 & 229 & 355 & 0.622 & 27.463 & 58.050 & -2.857 & -0.600 & 1.090 & 4 & 0 \\
	Stock\_12 & 302 & 240 & 372 & 0.212 & 353.870 & 52.623 & 8.571 & -1.917 & 2.287 & 10 & 8 \\
	Theia\_585 & 302 & 240 & 372 & 0.334 & 175.317 & -60.768 & -5.163 & 0.028 & 1.130 & 4 & 0 \\
	\enddata
	\tablecomments{Age$_{\rm best}$ is the best-fit cluster age. Age$_{\rm lower}$ 
		and Age$_{\rm upper}$ are the lower and upper bounds of the $1\sigma$ confidence 
		interval on the age, as defined in Section \ref{clusterage}.
		$N_{\rm WD,exp}$ is the expected number of WDs from single-star evolution, and 
		$N_{\rm esc}$ is the number of recovered escaped WD candidates. The remaining
		cluster parameters---$A_V$ ($V$-band extinction in mag), R.A. and Decl. (
		equatorial coordinates in deg), $\mu_{\alpha}\cos\delta$ and $\mu_{\delta}$
		(proper-motion components in mas\,yr$^{-1}$), and $\varpi$ (parallax
		in mas)---are taken from the H\&R24 OC catalog. (This table is 
		available in its entirety in machine-readable form.)}
\end{deluxetable*}

\subsection{Main Results}
Our initial kinematic search for escaped WDs from OCs using the best-fit cluster ages derived in Section~\ref{clusterage} yielded a raw sample of approximately 7,600 candidates. To filter out unassociated background field contamination from this catalog, we exploit WD cooling ages as an astrophysical constraint. We used the \texttt{WD\_models} Python code\footnote{\url{https://github.com/SihaoCheng/WD_models}} to derive the WD parameters, including masses, effective temperatures and cooling ages based on the cooling tracks from \citet{Bedard2020}. Specifically, we adopt the helium-atmosphere (DB) cooling age (age\_cool\_He) as a conservative lower limit. For a given effective temperature and surface gravity, helium-atmosphere WDs typically cool faster than their hydrogen-atmosphere counterparts, thereby providing the shortest possible cooling age. We use this helium cooling age as a selection criterion: if even this minimum plausible age exceeds the cluster lifetime, the source is rejected as a likely 
field interloper. This constraint is enforced quantitatively by requiring:	\begin{equation}
	\texttt{age\_cool\_He} \le \texttt{up\_age} + 3\,\bigl(\texttt{up\_age} - 
	\texttt{low\_age}\bigr),
\end{equation}
where \texttt{up\_age} and \texttt{low\_age} represent the upper and lower $1\sigma$ bounds of the cluster age obtained from isochrone fitting (Section~\ref{clusterage}). This cluster-specific threshold removes sources whose minimum plausible cooling 
age exceeds the cluster's upper age bound by more than three times its 
$1\sigma$ width. After applying this cooling-age consistency cut, our sample 
comprises 3,863 escaped WD candidates. 

To further account for the measurement uncertainties in Gaia astrometry, which could still introduce significant contamination among these age-consistent sources, we perform a Monte Carlo (MC) analysis for each candidate. We generate 500 MC realizations by sampling from Gaussian distributions centered on the observed parallax and proper motions, incorporating their associated uncertainties. The full kinematic escape evaluation procedure is executed independently for each individual realization. Ultimately, the final escape probability, $P_{\rm esc}$, is defined as the fraction of these 500 simulated trials that successfully satisfy our dynamic escape criteria.

\begin{figure*}[htbp]
	\centering
	\includegraphics[width=\textwidth]{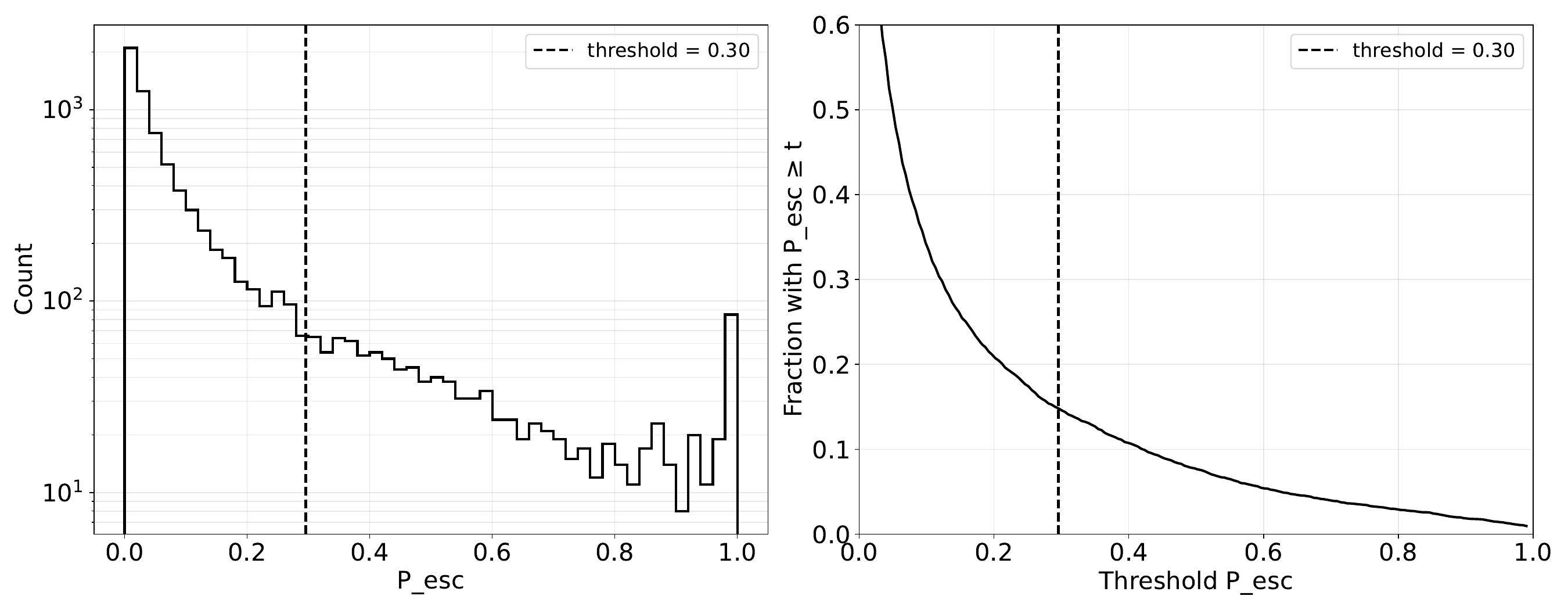}
	\caption{The determination of the $P_{\rm esc}$ threshold based on the differential distribution of candidates with a logarithmic y-axis (left) and their complementary cumulative fraction as a function of the probability threshold (right). The dashed vertical line indicates the data-driven inflection point at $t = 0.30$.}
	\label{fig:pesc_threshold}
\end{figure*}

\begin{figure*}
	\centering
	\includegraphics[width=\textwidth]{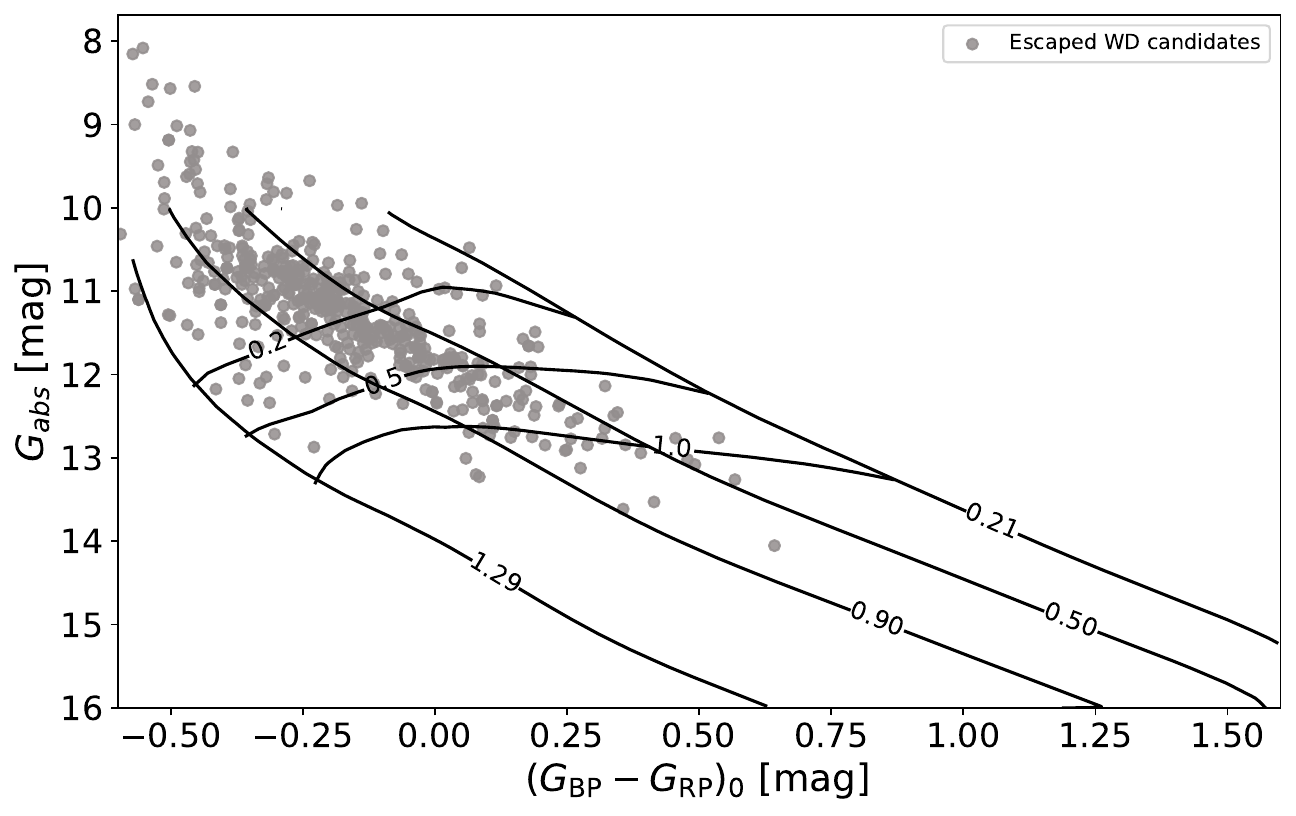}
	\caption{CMD of  escaped WD candidates. Gray circles represent the escaped WD candidates. Theoretical cooling tracks for DA WDs are from \citet{Bedard2020}, with nearly vertical lines showing constant-mass sequences (labeled in $M_\odot$) and nearly horizontal lines indicating constant cooling ages (labeled in Gyr).}
	\label{fig:fig1}
\end{figure*}

To set a data-driven threshold for $P_{\rm esc}$, we examine how the 
sample size varies with the probability cutoff $t$. We compute the fraction of 
	candidates with $P_{\rm esc} \ge t$, which decreases monotonically from 1 to 0 
	as $t$ increases (Figure~\ref{fig:pesc_threshold}, right). Both this cumulative 
	fraction and the differential distribution (left) show a clear shoulder near 
	$t = 0.3$, marking the transition from the steeply declining, 
	noise-dominated regime at low $P_{\rm esc}$ to the flatter regime populated by 
	plausible escapees. To locate this transition objectively, we compute the second 
	derivative of the log source count, $d^2 \log N(t)/dt^2$, whose maximum 
	identifies the point of strongest curvature (the ``elbow''). Below this elbow, 
	raising the threshold mainly removes background contaminants; above it the curve 
	flattens, since the surviving sources are predominantly genuine candidates that 
	would only be gradually eroded by a stricter cut. The second derivative peaks at 
	$t = 0.3$, so we adopt $P_{\rm esc} \ge 0.3$ as our escape threshold, balancing 
	completeness against purity. This yields a final sample of 476 WDs associated 
	with 175 OCs. Figure~\ref{fig:fig1} shows their distribution on the CMD, where 
	gray circles mark the escaped WD candidates.

The parameters of these 476 escaped WD candidates are listed in Table~\ref{tab:escaped_WDs}. We note that some candidates in this final catalog 
	are kinematically associated with more than one cluster. This arises when the 
	backward kinematic tracing of a WD's trajectory happens to align with the past 
	positions of several distinct clusters. Rather than indicating genuine 
	membership in multiple clusters, such cases reflect the kinematic degeneracy of 
	the method, and suggest that a fraction of these associations may be chance 
	alignments rather than true ones.


\begin{deluxetable*}{lrrrrcrrrrrr}
	\tablecaption{Parameters of the 476 escaped WD candidates. \label{tab:escaped_WDs}}
	\tablewidth{0pt}
	\tablehead{
		\colhead{Gaia DR3 ID} & \colhead{$d_{\rm present}$} & \colhead{$d_{\rm min}$} & \colhead{$\Delta v_{\rm 3D}$} & \colhead{$t_{\rm esc}$} & \colhead{Cluster} & \colhead{$p_{\rm esc}$} & \colhead{$\tau_{\rm cool}$} & \colhead{$\tau_{\rm prog}$} & \colhead{$M_{\rm WD}$} & \colhead{$M_{\rm init}$} & \colhead{label} \\
		\colhead{} & \colhead{(pc)} & \colhead{(pc)} & \colhead{(km\,s$^{-1}$)} & \colhead{(Myr)} & \colhead{} & \colhead{} & \colhead{(Gyr)} & \colhead{(Gyr)} & \colhead{($M_\odot$)} & \colhead{($M_\odot$)} & \colhead{}
	}
	\startdata
	2066386870087740288 & 146.331 & 5.730 & 2.753 & -51.928 & ASCC\_101 & 0.624 & 0.116 & 0.081 & 1.081 & 5.803 & 2 \\
	2087501204194756352 & 171.445 & 7.383 & 4.460 & -37.551 & ASCC\_101 & 0.788 & 0.219 & 1.715 & 0.701 & 1.886 & 3 \\
	2072866017242421888 & 61.037 & 14.078 & 4.758 & -12.206 & ASCC\_101 & 0.328 & 0.065 & 0.315 & 0.848 & 3.427 & 2 \\
	2020728683375177344 & 108.349 & 1.437 & 3.677 & -28.812 & ASCC\_101 & 0.722 & 0.165 & 0.075 & 1.096 & 6.006 & 2 \\
	2036061205536029952 & 79.356 & 1.757 & 4.268 & -18.175 & ASCC\_101 & 1.000 & \nodata & \nodata & \nodata & \nodata & \nodata \\
	\multicolumn{12}{c}{$\cdots$} \\
	1968733744872205312 & 195.764 & 12.784 & 3.353 & -56.971 & Stock\_12 & 0.370 & 0.044 & 0.248 & 0.906 & 3.738 & 2 \\
	2031057358175294720 & 318.489 & 8.395 & 2.553 & -121.937 & Stock\_12 & 0.640 & 0.069 & 1.979 & 0.676 & 1.736 & 3 \\
	1364054915694481408 & 344.396 & 5.906 & 4.968 & -67.773 & Stock\_12 & 0.330 & 0.216 & 0.353 & 0.822 & 3.291 & 2 \\
	5966130760147250304 & 521.072 & 5.164 & 4.730 & -107.713 & Stock\_12 & 0.756 & 0.086 & 2.791 & 0.642 & 1.536 & 3 \\
	1228266814506156928 & 460.209 & 8.946 & 3.835 & -117.318 & Stock\_12 & 0.914 & 0.268 & 3.191 & 0.631 & 1.468 & 3 \\
	\enddata
	\tablecomments{$d_{\rm present}$ is the present-day projected distance from the cluster center. $d_{\rm min}$ and $\Delta v_{\rm 3D}$ are the minimum approach distance and relative velocity at closest approach during back-tracing. $t_{\rm esc}$ is the estimated escape time. $p_{\rm esc}$ is the escape probability from Monte Carlo simulations. $\tau_{\rm cool}$ is the WD cooling age derived from H-atmosphere models. $\tau_{\rm prog}$ is the progenitor lifetime. $M_{\rm WD}$ is the WD mass. $M_{\rm init}$ is the initial progenitor mass. The label column classifies each candidate by comparing its single-star total age (cooling age plus progenitor lifetime) to the host cluster age: label~0 marks low-mass ($M_{\rm WD}<0.53\,M_\odot$) systems flagged as likely binary products; label~1 marks objects significantly younger than the host cluster; label~2 marks objects whose total age is consistent with the cluster, constituting the high-confidence sample of genuine escapees; and label~3 marks objects significantly older than the host cluster. (This table is available in its entirety in machine-readable form.)}
\end{deluxetable*}

\section{Discussion} \label{sec4}

\subsection{Contamination Assessment}

It is important to note that, although this sample has passed both the cooling-age consistency filtering and the kinematic MC screening, these criteria establish each candidate as a statistically  escapee rather than a 
confirmed member. As demonstrated by \citet{Miller2023} in their study of the Hyades star cluster, the inherent interloper fraction is a strong function of the WD mass distribution; even high-mass candidates ($1.1\,M_{\odot}$) can exhibit a $\sim 50\%$ chance of being interlopers, with this contamination rate expected to increase for more typical WD masses. 

To estimate the contamination fraction arising from chance kinematic 
	coincidences in our candidate sample, we carried out a MC null test 
	based on control (mock) clusters. The idea is to build a set of mock clusters 
	whose positions and velocities are randomly offset from those of the real 
	clusters, which removes their physical association with any escaped WDs while 
	preserving the local Galactic environment---and hence the surrounding field-WD 
	density and kinematics. We then re-applied the full escape-detection pipeline to 
	these mock clusters. By construction the mock clusters have no real WD escapees, 
	so any source that passes the selection should mostly be a field star that 
	aligns with a mock cluster by chance. Comparing the number of such spurious 
	detections with the real candidate count then gives an empirical estimate of the 
	contamination rate.

In practice, for each of the OCs ($N_{\rm cl} = 175$ ) that host at least one  candidate, we generated $N_{\rm ctrl}=100$ mock clusters. The centre of each mock cluster was displaced from the true cluster centre by a random offset uniformly distributed in volume between $100$ and $500\,{\rm pc}$, and its bulk velocity was perturbed by a three-dimensional Gaussian with $\sigma_v = 10\,{\rm km\,s^{-1}}$, while the cluster age was kept unchanged. This design approximately preserves the local Galactic environment while removing any genuine physical association with the WDs. We then re-ran the identical escape-detection pipeline on the full WD catalogue against every mock cluster, including the 500-realization Monte Carlo perturbation of parallax and proper motion for each WD, recomputing $P_{\rm WD}$ and the escape probability $P_{\rm esc}$, and recorded all sources that satisfied the same selection thresholds.

For a given set of thresholds $(P_{\rm WD}\ge a,\; P_{\rm esc}\ge b)$, the global contamination fraction is defined as $C(a,b) = \langle N_{\rm fake}(a,b) \rangle \,/\, N_{\rm real}(a,b)$, where $N_{\rm real}(a,b)$ is the number of real candidates passing the thresholds, and $\langle N_{\rm fake}(a,b) \rangle = \sum_{i=1}^{N_{\rm cl}} \langle N_{\rm fake}^{(i)} \rangle$ is the total expected number of mock candidates, with $\langle N_{\rm fake}^{(i)} \rangle$ being the arithmetic mean over the $N_{\rm ctrl}$ mock trials for cluster $i$. To estimate the $1\sigma$ uncertainty on $C$, we adopt the more conservative of two error estimates. First, we compute the standard error of the mean for each cluster as $\sigma_i = \operatorname{std}_k(N_{{\rm fake},ik}) \,/\, \sqrt{N_{\rm ctrl}}$, and then propagate these into an empirical uncertainty $\sigma_C^{\rm emp} = \sqrt{\sum_{i=1}^{N_{\rm cl}} \sigma_i^2} \,/\, N_{\rm real}$. Second, a Poisson-based uncertainty is given by $\sigma_C^{\rm Pois} = \sqrt{\langle N_{\rm fake} \rangle} \,/\, N_{\rm real}$. We then take $\sigma_C = \max(\sigma_C^{\rm emp},\; \sigma_C^{\rm Pois})$ as the final conservative $1\sigma$ error on the contamination fraction. For our adopted selection criteria of $P_{\rm WD} \ge 0.9$ and $P_{\rm esc} \ge 0.3$, the resulting contamination fraction is $87.6\% \pm 4.3\%$.

\subsection{The High-confidence Sample}

When tracing escaped WDs on timescales of tens to hundreds of Myr, 
high contamination rates are inevitable due to intrinsic physical challenges: 
the extremely high stellar density of the Galactic disk, severe kinematic 
overlap, the intrinsic faintness of WDs, and limited astrometric 
precision. To reliably distinguish genuine members from contaminants, 
we classify the sample by comparing the total age of each WD with 
the cluster age.

To calculate the WD progenitor ages, we first derived their initial masses 
from the estimated WD masses utilizing the recent initial-final mass relation 
(IFMR) provided by \citet{Miller2026}. Subsequently, we constructed a dense 
grid of PARSEC isochrones to establish a robust initial mass-lifetime relation. 
For this grid, we adopted a fixed solar-like metallicity of $Z=0.015$ and 
generated isochrones over an age range from $\log(\mathrm{Age}/\mathrm{yr})=7.00$ 
to $10.35$ with a fine step of $0.01$~dex. Within each isochrone, we specifically 
identified the lowest-mass stars that have reached the early asymptotic giant 
branch (EAGB) stage, recording their initial masses to represent the exact 
progenitor lifetime at that given age. Finally, we interpolated this finely 
sampled mass-lifetime relation to infer the progenitor lifetime for each 
specific initial mass. The total age of each WD candidate was then obtained 
by summing its cooling age and progenitor age. For a more stringent constraint 
on the sample precision, cooling ages are uniformly derived using DA WD cooling models because the vast majority of WDs in the local 
sample are of DA type \citep{Brien2023}, providing more consistent age constraints than DB models.

We classify each WD candidate by comparing its total age with the cluster's age range. For each cluster, we define a young threshold as $\mathrm{low\_age} - N \times (\mathrm{up\_age} - \mathrm{low\_age})$ and an old threshold as $\mathrm{up\_age} + N \times (\mathrm{up\_age} - \mathrm{low\_age})$, where $\mathrm{low\_age}$ and $\mathrm{up\_age}$ are the lower and upper bounds of the cluster age adopted from the literature, and $N$ is a scaling factor (we adopt $N=3$, treating the cluster age range as an approximate $\pm 1\sigma$ boundary, so that a factor of three corresponds to a conservative $3\sigma$-equivalent exclusion zone). If the WD total age falls below the young threshold, it is considered significantly younger than the cluster and labeled as label~1; if it exceeds the old threshold, it is significantly older and labeled as label~3.
WDs with total ages lying between these two boundaries are consistent with the cluster age spread and are labeled as label~2. Additionally, WDs with masses below $\sim 0.5\,M_\odot$ are expected to helium core WD. According to single-star evolution, such low-mass WDs cannot form within a Hubble time and are generally the products of binary interactions. Since the IFMR of \citet{Miller2026} adopts a lower mass limit of $0.53\,M_\odot$, we classify all escaped WD candidates with $M_{\mathrm{WD}} < 0.53\,M_\odot$ as binary-evolution products and assign them label~0. The length of the progenitor lookback time may affect the 
accuracy of identifying escaped stars. In our analysis, the 97.5\% percentile 
of the lookback time for the 476 sample stars is only 302.2~Myr, indicating 
that the overall lookback timescale is not large; therefore, no additional 
restriction is imposed on the lookback time.

Applying this classification to our sample of 476 WD candidates, we find 90 low-mass binary products (label~0), 1 source that is significantly younger than its host cluster (label~1), 109 WDs whose total ages are consistent with the cluster age spread (label~2), and 252 significantly older objects (label~3). An additional 24 sources fall outside the applicable range of the WD models and are left unclassified (NaN). The corresponding statistics are listed in Table~\ref{tab:escaped_WDs}. The 109 label~2 WDs constitute our high-confidence member sample.

\subsection{WDs Formed Through Binary Evolution}

Binary interactions and internal physical mechanisms can alter the apparent evolutionary timescale of a WD. On one hand, mass transfer or CE evolution can accelerate the evolution of a stellar progenitor by prematurely stripping the envelope of a giant star, leading to the formation of a WD on a much shorter timescale, such as in the case of ELM WDs \citep[e.g.,][]{Kilic2011}. Furthermore, the violent ejection of a CE typically leaves the resulting WD with a substantially thinner residual hydrogen layer ($M_H$), which in turn accelerates its subsequent cooling rate \citep{althaus2026extrememasslosscommon}. Conversely, mechanisms such as binary mergers or internal physical phase transitions can introduce a substantial cooling delay. Particularly for high-mass WDs, the release of latent heat and gravitational potential energy from elemental settling causes an anomalous cooling pause on the Q branch of the CMD \citep[e.g.,][]{Cheng2019}. This additional energy effectively slows the apparent cooling rate, making the remnant appear significantly younger than its actual true age. Consequently, WDs that are kinematically associated with the cluster based on astrometry but possess derived total ages deviating from the cluster age may be the products of complex binary evolutionary channels.  Thus, sources labelled 0, 1, and 3 may imply higher field-star contamination and also include binary evolution products.

According to binary population synthesis models, about 10\%--30\% of normal and intermediate-mass ($M < 0.9 M_\odot$) single WDs are formed through binary mergers \citep{Temmink2020}, with the majority of them involving the descendants of mergers between post-main-sequence and main-sequence stars. For massive WDs ($M \geq 0.9 M_\odot$), where the merger channel is more prevalent, the expected merger fraction reaches 30\%--45\% \citep{Temmink2020}.
%
 Close double white dwarf (DWD) binaries are estimated to account for $\sim$9\% of the WD population based on the latest SDSS DR19 results \citep{Adamane2025}. We compare the fraction of age-anomalous WDs in our sample (those classified as label~0, 1, and~3) with these expectations in different mass intervals. For $0.53$--$0.9\,M_\odot$, the age‑anomalous fraction of $76.21\%$ far exceeds the $10\%$--$30\%$ merger contribution predicted by population synthesis \citep{Temmink2020}, whereas for $>0.9\,M_\odot$ the corresponding fraction of 32\% falls within the expected $30\%$--$45\%$. This contrast suggests that the low-mass age-anomalous WDs might be more influenced by field-star contamination, while the high-mass ones might be plausibly explained by binary evolution.

This high age anomaly rate is not unprecedented. Using Gaia DR3 astrometry, \citet{Yan2026} found 439 WD candidates across 117 OCs and also reported a large fraction of WDs with derived total ages significantly surpassing their respective cluster ages (244 WD candidates exceed the age of their parent clusters). Beyond interloper contamination, however,, such age discrepancies may also expose  limitations within current theoretical frameworks. For instance, the standard WD cooling models adopted here might not fully capture unconventional internal crystallization or complex envelope physics. Furthermore, uncertainties in macroscopic stellar evolution---particularly regarding mass-loss rates and common-envelope ejection efficiencies---can systematically bias progenitor age estimates. Follow-up spectroscopy will therefore be valuable: accurately determining  the atmospheric parameters and chemical compositions of these candidates would provide empirical constraints to help separate genuine cluster members from field interlopers and to refine current evolutionary models. 

\subsection{Correlations with Cluster Properties}
The observed number of WDs in OCs is substantially lower than theoretical 
predictions, and one important reason is that a large fraction of WDs escape 
from their parent clusters over time. In this work, we systematically searched 
for escaped WDs originating from OCs using a kinematic cross-matching method. 
To evaluate the performance of this approach and understand what governs the 
number of recovered escaped WDs, we examine how $N_{\rm escaped}$ and the 
escape fraction $N_{\rm escaped}/\langle N_{\rm WD} \rangle_{\rm int}$ depend 
on three fundamental cluster properties: heliocentric distance, age, and 
mass.

To compute the escape fraction, we first estimate the theoretically expected 
number of WDs produced through single-star evolution in each OC following the 
method of \citet{Richer2021}. We adopt a Kroupa IMF $dN/dM$ 
\citep{Kroupa2001, Kroupa2002} and integrate from the current WD-producing 
mass $M_{\rm init,\,WD}$ up to an upper mass limit of $8\,M_\odot$, normalized 
by the observed number of bright MS stars. Specifically, we select the 
brightest one-third of MS stars up to the turnoff. The normalization factor $N_1$ is the observed count of stars in this 
bright subsample, and the turnoff point is identified as the highest effective 
temperature reached along the main sequence. The expected WD number is then 
given by:
\begin{equation}
	\langle N_{\rm WD} \rangle = N_1
	\left[ \int_{M_{\rm init,3}}^{M_{\rm init,\,TO}} \frac{dN}{dM}\, dM 
	\right]^{-1}
	\int_{M_{\rm init,\,WD}}^{8\,M_\odot} \frac{dN}{dM}\, dM,
	\label{eq:NWD}
\end{equation}
\noindent where $M_{\rm init,3}$ is the initial mass corresponding to the 
faintest star in the bright subsample and $M_{\rm init,\,TO}$ is the turnoff 
mass. Throughout this section, we use the integer-rounded value 
$\langle N_{\rm WD} \rangle_{\rm int}$ as the denominator of the escape 
fraction, and restrict the analysis to the 175 clusters for which at least 
one escaped WD candidate was identified. 

\begin{figure*}
	\centering
	\includegraphics[width=\textwidth]{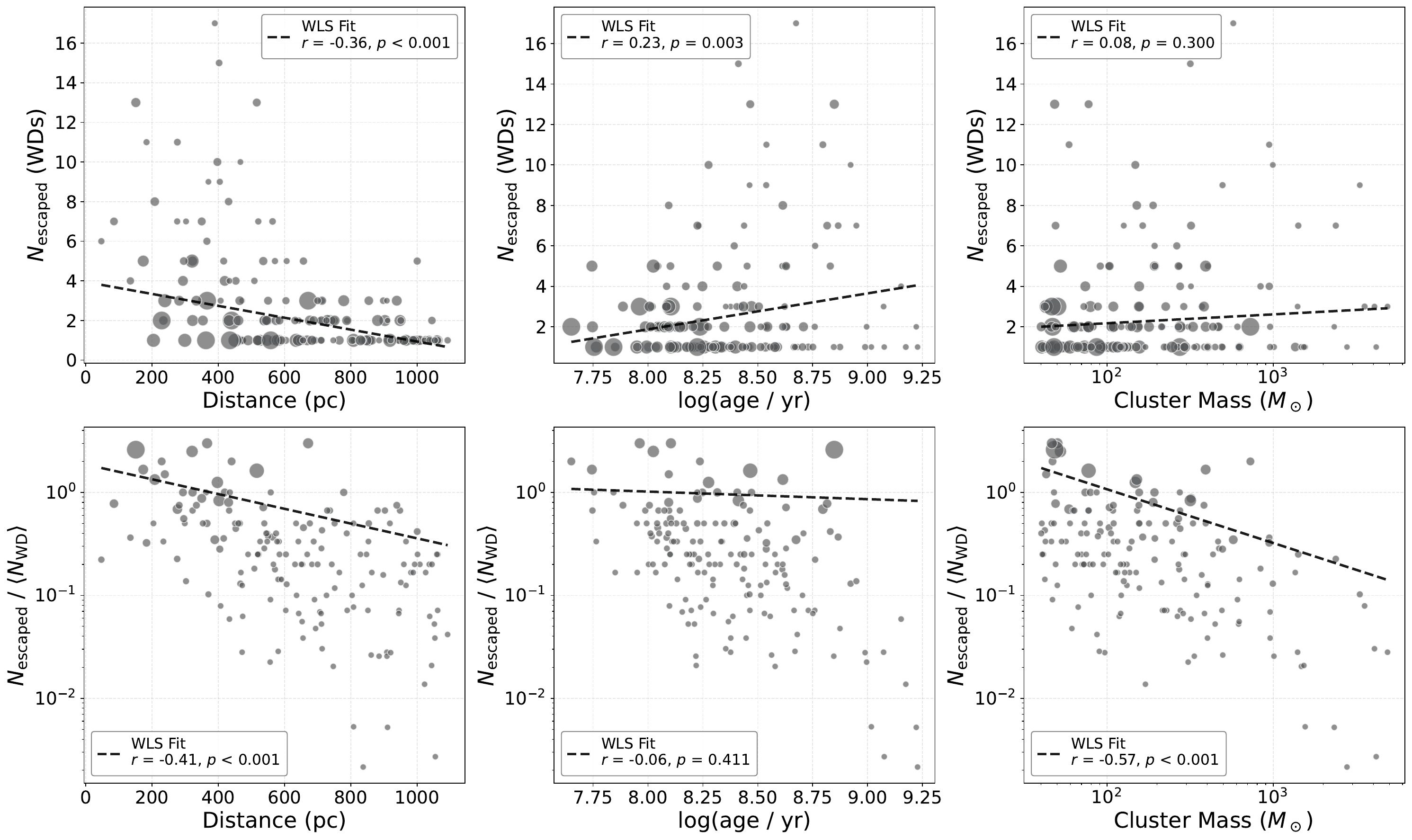}
   \caption{Escaped WD count $N_{\rm escaped}$ (upper panels) and escape fraction 
	$N_{\rm escaped}/\langle N_{\rm WD} \rangle_{\rm int}$ (lower panels) as a 
	function of heliocentric distance (left), cluster age (middle), and 
	present-day mass (right) for the 175 OCs. The dashed lines show 
	the best-fit linear regression, performed in log-space for the escape fraction 
	panels. The Pearson correlation coefficient $r$ and the corresponding $p$-value 
	are indicated in each panel. The size of each point is proportional to the WLS fit weight derived from $N_{\rm escaped}$ and $N_{\rm escaped}$, with larger points carrying greater statistical weight in the fit. }
	\label{fig:3}
\end{figure*}

Figure~\ref{fig:3} shows the escaped WD count $N_{\rm escaped}$ (upper 
panels) and the escape fraction $N_{\rm escaped}/\langle N_{\rm WD}\rangle_{\rm int}$ 
(lower panels) as functions of heliocentric distance (left), cluster age 
(middle), and cluster mass (right) for the 175 OCs in our sample. The 
dashed lines show the best-fit weighted least-squares (WLS) relations, 
performed in log-space for the escape-fraction panels, and each panel lists 
the Pearson coefficient $r$ together with its $p$-value.

In absolute terms, $N_{\rm escaped}$ is anticorrelated with 
	distance ($r = -0.36$, $p < 0.001$), which may reflect Gaia's magnitude-limited 
	completeness: more distant escaped WDs tend to be fainter and more dispersed, and 
	so may be more likely to fall below the detection threshold. A weak positive 
	correlation with age ($r = 0.23$, $p = 0.003$) is also seen, which could be 
	broadly consistent with a gradual build-up of the WD population, since older 
	clusters have had more time to produce WDs. No significant correlation with 
	cluster mass is found ($r = 0.08$, $p = 0.300$); this might arise from two 
	competing tendencies, as massive clusters may host larger intrinsic WD populations 
	but could lose a smaller fraction of them, which would tend to leave little net 
	trend in the absolute count.

The escape fraction, which normalizes the recovered count by the predicted 
	intrinsic WD population, might help to isolate any dynamical signal somewhat more 
	cleanly. It too appears anticorrelated with distance ($r = -0.41$, $p < 0.001$), 
	perhaps because the denominator is a model prediction, so that incompleteness may 
	suppress mainly the numerator and could tend to drive the ratio down with distance. 
	The escape fraction shows almost no dependence on age ($r = -0.06$, $p = 0.411$). 
	This is hard to square with a picture of slow, steady evaporation: if white dwarfs 
	were gradually shed over time, older clusters should have lost a larger fraction of 
	them. A more likely explanation may be that the losses happen mainly when the WDs first form---for example, through natal kicks, which give the WD a 
	velocity boost at formation. Because these clusters have fairly low escape 
	velocities, even a modest kick could be enough to eject them. The escape fraction 
	also appears to be more clearly anticorrelated with cluster mass ($r = -0.57$, 
	$p < 0.001$): more massive clusters, with deeper potential wells, might be expected to retain a larger fraction of their WDs, while lower-mass clusters might tend to lose more of theirs.

\section{Conclusion}\label{sec5}

In this work, we present a systematic search for escaped WDs originating from
OCs using astrometric and photometric data from Gaia DR3. We identify
approximately 7,600 escaped WD candidates. After excluding sources whose
cooling ages significantly exceed the ages of their host clusters, and
incorporating MC simulations to account for astrometric
uncertainties, we obtain a final sample of 476 escaped WD candidates in 175 OCs.

To quantify the contamination fraction arising from chance kinematic coincidences,
we performed a control-field MC test using mock clusters whose
positions and velocities were randomly perturbed to break any physical association
with WDs while preserving the local Galactic environment.
Re-running the full escape-detection pipeline on these mock clusters yielded an
expected number of spurious candidates, comparing this number with the real candidate count, defines the global contamination fraction.
Adopting a conservative error estimate, our selection criteria
($P_{\rm WD} \ge 0.9$, $P_{\rm esc} \ge 0.3$) result in a contamination rate of
$87.6\% \pm 4.3\%.$

To reduce contamination in our kinematic sample, we estimated single-star 
	total ages for the 476 escaped WD candidates and compared them to the host 
	cluster ages. We find 109 high-confidence genuine escapees (label~2), while the 
	remainder show age anomalies or low masses indicative of binary products. The 
	age-anomalous fraction reaches $76\%$ for $0.53$--$0.9\,M_\odot$, far exceeding the 
	$10\%$--$30\%$ predicted by binary population synthesis, whereas the $32\%$ fraction 
	at $>0.9\,M_\odot$ remains consistent with the expected $30\%$--$45\%$.

Furthermore, we investigated the factors governing the observational 
	recovery of escaped WDs. The absolute number of detected escapees per cluster 
	appears to be limited mainly by Gaia's distance-dependent photometric 
	incompleteness, with a weak positive trend with cluster age and no significant 
	dependence on cluster mass. To probe the intrinsic escape efficiency, we 
	normalized these counts by the expected WD population from single-star evolution. 
	The resulting escape fraction remains anticorrelated with distance but shows 
	almost no dependence on age---possibly disfavouring gradual secular evaporation 
	in favour of WD loss near the epoch of formation, such as natal kicks. It also 
	shows a clearer anticorrelation with cluster mass, plausibly because more 
	massive clusters, with deeper potential wells, may retain a larger fraction of 
	their WDs.

\begin{acknowledgments}
	C.Y.L. and H.H.Y. acknowledge support from the National Natural Science Foundation of China (NSFC) through grant 12233013. D.R.M acknowledges support from the Natural Sciences and 
	Engineering Research Council of Canada Discovery Grants 
	DG-RGPIN-2022-03051 and DG-RGPIN-2023-04486. H.H.Y. thanks the China Association for Science and Technology (CAST) Young Elite Scientists Sponsorship Program for support. J.K.Z acknowledges support from the NSFC (Grant No. 12273055). J.C.G. acknowledges support from the Young Scholar Program of the Beijing Academy of Science and Technology (Grant No. 25CE-YS-02) and the NSFC (Grant No. 12203006).
\end{acknowledgments}

\bibliography{sample701}{}
\bibliographystyle{aasjournalv7}



\end{document}